\ifx\mnmacrosloaded\undefined 
\input mn.tex\fi
\input psfig.sty


\def\eg{{e.g.~}}
\def\ie{{i.e.~}}
\def\etal{et~al.~}
\def\eV{e\kern-.15em V}                 
\def\keV{ke\kern-.15em V}                
\def\nh{$N_{\rm H}$}
\def\nhint{$N_{\rm Hint}$}

\def\Ha{H$\alpha$}
\def\Hb{H$\beta$}
\def\oiii{[O{\sc iii}]}

\def\feii{Fe{\sc ii}}

\def\aox{$\alpha_{\rm ox}$}
\def\aro{$\alpha_{\rm ro}$}
\def\arx{$\alpha_{\rm rx}$}

\def\ax{$\alpha_{\rm x}$}
\def\aopt{$\alpha_{\rm opt}$}

\def\nhgal{$N_{\rm HGal}$}
\def\nhint{$N_{\rm Hint}$}
\def\kms{km~s$^{-1}$}

%

\newif\ifAMStwofonts

\ifCUPmtplainloaded \else
  \NewTextAlphabet{textbfit} {cmbxti10} {}
  \NewTextAlphabet{textbfss} {cmssbx10} {}
  \NewMathAlphabet{mathbfit} {cmbxti10} {} 
  \NewMathAlphabet{mathbfss} {cmssbx10} {} 
  \ifAMStwofonts
    \NewSymbolFont{upmath} {eurm10}
    \NewSymbolFont{AMSa} {msam10}
    \NewMathSymbol{\upi}     {0}{upmath}{19}
    \NewMathSymbol{\umu}     {0}{upmath}{16}
    \NewMathSymbol{\upartial}{0}{upmath}{40}
    \NewMathSymbol{\leqslant}{3}{AMSa}{36}
    \NewMathSymbol{\geqslant}{3}{AMSa}{3E}

     \let\le=\leqslant
     \let\ge=\geqslant
  \else
    \def\umu{\mu}
    \def\upi{\pi}
    \def\upartial{\partial}
  \fi
\fi


\pageoffset{-2.5pc}{0pc}

\loadboldmathnames



\pagerange{000--000}    
\pubyear{0000}
\volume{000}

\begintopmatter  

\title{X-ray selected active galactic nuclei with red optical continua}
\author{E. M. Puchnarewicz and K.~O.~Mason}

\affiliation{Mullard Space Science  Laboratory, University College London,
Holmbury St. Mary, Dorking, Surrey RH5 6NT, UK.}

\shortauthor{E. M. Puchnarewicz \& K. O. Mason}
\shorttitle{Red AGN}



\abstract {We discuss the properties of X-ray selected `red' AGN from the RIXOS
sample. These are Seyfert 1 galaxies and quasars whose optical continua are
relatively soft, \ie with an energy index, \aopt$>$2. There are 14 objects in
the RIXOS sample which satisfy this criterion and they cover a range in
redshift from $z$=0.08 to 1.27. Of these, two have characteristics which
suggest that the continuum is intrinsically red, \ie an optical continuum which
does not appear to have been significantly  reddened by dust or to have
contaminating light  from the host galaxy.  A further three objects show
evidence of being absorbed by cold gas and dust with columns of up to
$\sim10^{22}$ cm$^{-2}$. The data are inconclusive on the remaining AGN.

}


\maketitle  

\section{Introduction}

The `big blue bump' (BBB), a steep rise towards high frequencies in the
optical/UV range, has been observed in many non-blazar AGN where it dominates
the bolometric luminosity (\eg Edelson \&\ Malkan 1986; Elvis \etal 1994a). 
The apparent ubiquity of BBBs in AGN has justified the use of optical and
UV-excess selection methods as an effective way of defining and identifying 
samples of AGN.  However, deep optical observations of the blank fields around
previously unidentified radio sources, have revealed a number of  `red' quasars
(also termed `optically-quiet' and `optically-dull': see \eg Rieke, Lebofsky \&
Kinman 1979; Ledden \& O'Dell 1983;  Bregman \etal 1985; Ulvestaad \& Antonucci
1986; Elvis \etal 1994b; Kollgard \etal 1995). These quasars have unusually
high radio-to-optical ratios and their colours suggest a steep, soft,
IR-to-optical continuum slope.

Detailed studies of several of these objects have been made and in all cases 
the underlying cause of the red continuum is likely to be reddening by dust
(\eg Elvis \etal 1994b; Kollgard \etal 1995). Ledden \& O'Dell (1983) found
that optically-quiet radio sources also tended to be weak in X-rays and
interpreted this as being due to the effects of absorption as well. These
objects can provide us with some idea of the effects of dust on the overall
quasar population, although even these may only represent a small minority
of the reddened AGN population. This situation has major consequences for our
interpretation of quasar properties, yet is notoriously difficult to tackle
because of the problems in finding an unambiguous measurement of the amount of
dust present (see \eg Grandi 1983), and thus deriving the nature of the
intrinsic AGN continuum.

A red continuum may have other explanations besides the presence of dust
however. Two AGN with very strong EUV emission, RE~J1034+396 and  RE~J1237+264
(Puchnarewicz \etal\ 1995; Brandt, Pounds \& Fink 1995) have optical/UV
continua which rise steeply to the red, yet their strong EUV/soft X-ray
emission and low Balmer decrements argue against absorption. Similar properties
are observed in the IR-selected IRAS13349+2438, although a {\sl warm} (\ie
ionized), dusty absorber has been suggested for this AGN (Wills \etal 1992;
Brandt, Fabian \& Pounds 1996) and similar dusty warm absorbers have  been
inferred in the Seyfert galaxies MCG~6-30-15 (Reynolds \etal 1997) and NGC~3227
(Komossa \& Fink 1997). Studies of RE~J1034+396 have shown that the optical/UV
continuum contained neither significant  emission from the BBB nor from the
host galaxy (Puchnarewicz \etal 1995), while the  2-10~\keV\ {\sl ASCA}
spectrum (Pounds, Done \& Osborne 1996) and {\sl IRAS} 12-100$\mu$m colours are
consistent with an extrapolation of the optical/UV slope to higher and lower
energies. In addition, the optical spectrum does not exhibit  any linear
polarization, with an upper limit of  $\sim$0.3-0.4 per cent (Breeveld \&
Puchnarewicz 1997). All of this evidence points to an {\sl intrinsically} red
optical/UV slope in RE~J1034+396, which is part of an underlying power-law
component and extends from 100$\mu$m to 10~\keV.

Objects like these and the radio-selected red quasars have led authors to
comment on their similarity to blazars and speculate on a possible link between
these and non-blazar AGN (\eg Rieke \etal\ 1979; Bregman \etal 1985;
Puchnarewicz \etal 1995; Tananbaum \etal 1997). The existence of IR to X-ray
power-law continua in non-blazars has been disputed however; while observations
indicate that such a component may be common in AGN  (Malkan 1984; Elvis \etal
1986), it has been demonstrated that such effects may be mimicked by the sum of
several thermal components (\eg Barvainis 1993). If a power-law component could
be confirmed, it would have implications for non-thermal emission mechanisms in
AGN and for the relationship between AGN and blazars.

Both types of red AGN, \ie intrinsically-red objects and  dust-reddened
sources, will have been selected against when UV-excess techniques are used to
search for and identify AGN.  Previous identifications of red AGN have largely
been made by means of the optical identification of radio sources (\eg Rieke
\etal 1979; Bregman \etal 1985). Kollgard \etal (1995) used the {\sl ROSAT}
All-Sky Survey to search for their `optically-quiet' quasars, although
radio-loudness was a defining criterion in their study. In this paper, we have
used the X-ray (0.5-2~keV) selected RIXOS sample of AGN (Mason \etal, in
preparation)  to investigate the incidence of red AGN, independent of their
radio loudness. The properties of the RIXOS AGN sample as a whole are presented
in Puchnarewicz \etal (1996; the continua, hereafter Paper I) and Puchnarewicz
\etal (1997; emission lines, hereafter Paper II).

While selection in any restricted wavelength range is not perfect, the
advantages of using the RIXOS sample for the study of red AGN  are {\sl (1)}
that it reaches X-ray fluxes typically 10 times fainter than the EMSS (Gioia
\etal 1990; Stocke \etal 1991); and {\sl (2)} UV-excess was {\sl not} a
determining factor when an AGN identification was made. Thus these AGN can
suffer a greater degree of cold gas absorption before they become too faint to
detect, providing a wider range of absorbing columns,  and AGN with red optical
slopes would not have been discriminated against. This sample provides an
essential comparison between radio-selected and X-ray-selected red AGN. We
investigate the nature of any absorption and the possibility that the optical
slope may be a bare, intrinsically red power-law continuum with no BBB  (at
least down to the near-UV). Finally, the implications of these results on the
nature of AGN emission processes and on population studies of AGN are
discussed.  

\section{Data reduction}

\subsection{The RIXOS AGN and the red subsample}

The RIXOS sample (Mason \etal in preparation) is made up of objects discovered
serendipitously in medium deep (exposure $>$8~ksec), high Galactic latitude
($\vert b\vert>28^\circ$) pointed observations made with the {\sl ROSAT}
Position Sensitive Proportional Counter (PSPC; Pfefferman \etal 1986). Only
sources within 17~arcmin of the centre of the field and with a flux greater
than 3$\times10^{-14}$ erg cm$^{-2}$ sec$^{-1}$ in the {\sl ROSAT} `hard' band
(0.4-2.0~\keV) are used; optical identification of survey sources  is 94 per
cent complete to this flux level over 15 deg$^2$. This has produced a sample of
X-ray emitting AGN (\ie Seyfert 1s to 1.9s and quasars), selected irrespective
of the strength or shape of the optical/UV continuum. There are 160 objects in
this RIXOS subsample with unambiguous redshifted line emission and broad
permitted line widths [\ie with full widths at half maximum (FWHM) of
1000~\kms\ or more; see Paper II]. 

\beginfigure{1} 
\psfig{figure=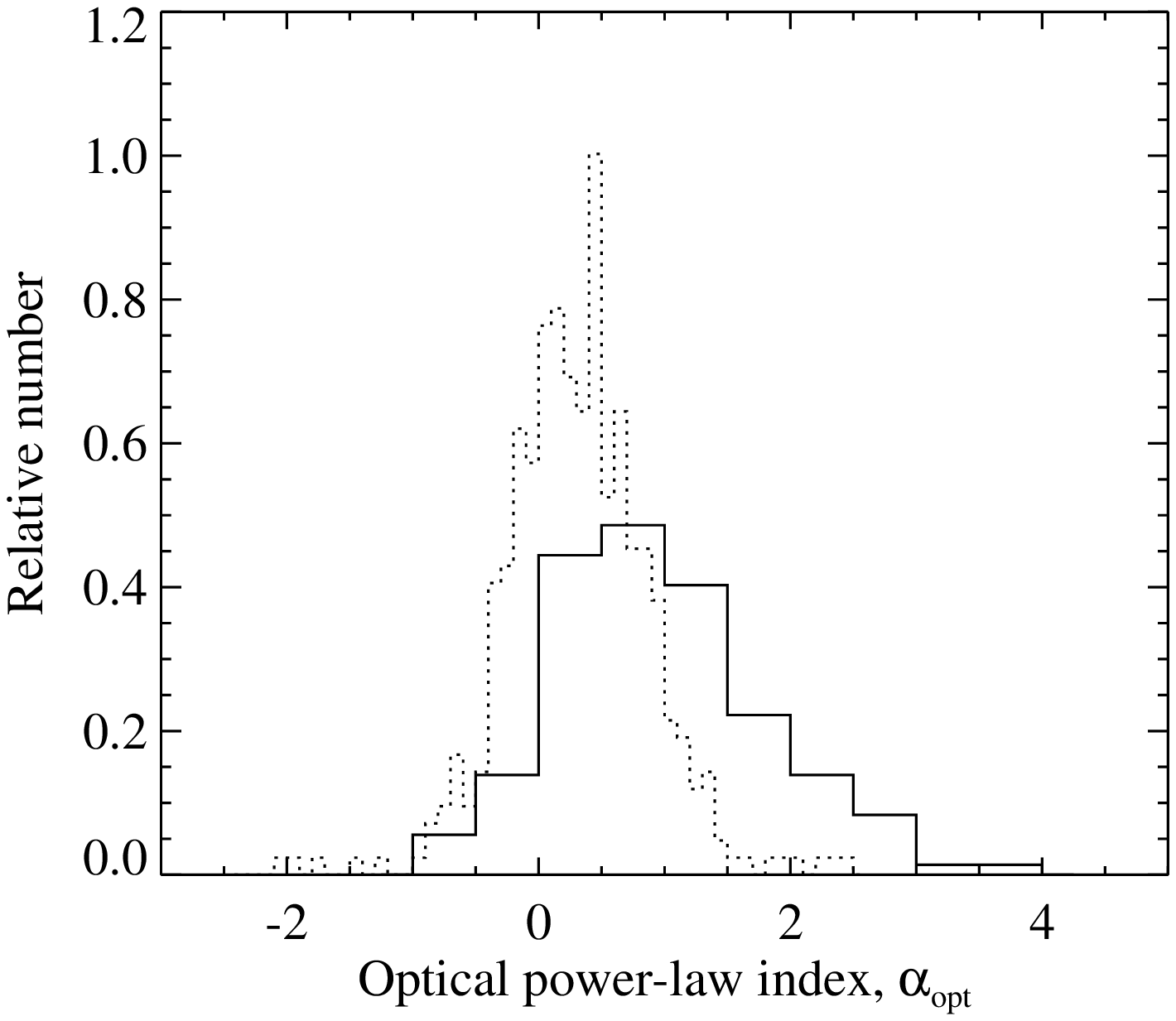,height=2.5in,width=3in,angle=0}
\caption{{\bf Figure 1.} The distribution of optical power-law indices, \aopt,
for the RIXOS AGN (plotted as a solid line) and compared with the bright
quasars from the Francis \etal (1991) sample (dotted line). Both distributions
have been normalized by area.}
\endfigure

\subsubsection{Measuring the optical continuum slope}

Optical spectral indices, \aopt, have been measured for 145 of these AGN by
fitting a simple power-law to the data, having first removed all absorption and
emission features and regions with a very low signal-to-noise ratio. For the
remaining 15 objects, either no spectrum was available [because the source had
been previously identified and the optical slope was not available in the
literature (6 sources)], or the spectrum was not taken with the slit at the
parallactic angle (9 sources). Although Balmer continuum and optical \feii\
emission can blend to produce a ``quasi''-continuum at around 3000~\AA\ (Wills,
Netzer \& Wills 1985) which may affect measurements of \aopt, this has been
reduced by removing these features and by using the  broadest wavelength range
available ($\sim$5000~\AA\ in the observer-frame). Full details of the optical
observations and derivations of the optical power-law slopes may be found in
Paper II.

Errors on the optical slopes are dominated by systematics and are difficult to
determine for individual spectra. We have assessed the typical error expected
on the slopes from the dispersion in the measured values for sources from the
parent sample which were observed more than once and estimate that the
uncertainty in \aopt\ is conservatively $\sim\pm$0.5.

\subsubsection{Red sample selection}

The distribution of \aopt\ for the RIXOS AGN is shown in Figure 1 (\aopt\ and
all indices $\alpha$ are defined such that $F_\nu\propto\nu^{-\alpha}$). The
mean \aopt\ is 0.9$\pm$0.1 (the error quoted is the error on the mean), which 
is softer than typical median values for UV-excess selected quasars where
\aopt$\sim$0.2-0.3 (\eg Neugebauer \etal 1987; Francis \etal 1991). The
distribution of the Francis \etal (1991) sample is compared with the RIXOS AGN
in Fig 1. This illustrates the tendency for the X-ray selected RIXOS objects
to have softer optical slopes than optically-selected AGN and suggests that a
significant population of `red AGN' may be missing from optically-selected
samples.


For the purposes of this study, we have (arbitrarily) chosen \aopt$\ge$2 as the
definition of a `red' AGN. This selects the 14 softest AGN from the RIXOS
sample; the softest 10 percent of objects. They are also very soft when
compared to optically-selected quasars, \eg in the Francis \etal (1991) sample
shown in Fig. 1, only 4 of the 419 quasars had \aopt$\ge$2. The RIXOS red AGN
source list including redshifts, J2000 positions, spectral V-band magnitudes
and \aopt\ are given in Table 1.

\begintable*{1}
\caption{{\bf Table 1.} RIXOS red AGN source list}
\halign{#\hfil                                                   
       &\hskip0.1truecm#\hfil                                    
        &\hskip0.1truecm\hfil#\hskip0.2truecm                    
        &\hskip0.2truecm#\hfil\hskip0.05truecm                   
       &\hskip0.05truecm#\hfil\hskip0.05truecm                   
       &\hskip0.2truecm\hfil#\hfil\hskip0.2truecm               
        &\hskip0.2truecm\hfil#\hskip0.2truecm                    
        &\hskip0.2truecm\hfil#\hskip0.2truecm                    
        &\hskip0.2truecm\hfil#\hskip0.2truecm                    
        &\hskip0.2truecm\hfil#                                   
\cr
\noalign{\bigskip}
    FID   &
    SNo   &
    \hfil{\sl z} \hfil   &
    \hfil{RA (J2000)} \hfil   &
    \hfil{Dec (J2000)} \hfil   &
    \hfil{m$_{\rm V}$} \hfil   &
    \hfil{\aopt}  &
    \hfil{\aox}   &
    \hfil{\aro}   &
    \hfil{\arx} 
 \cr
\noalign{\medskip}
122 & 13  & 0.358 & 16 30 55.10 & +78 11 02.0 & 20.6 & 3.1 & 1.0 & $<$0.4 & $<$0.5 \cr
122 & 21  & 0.376 & 16 34 27.80 & +78 10 03.0 & 20.8 & 2.3 & 1.0 & $<$0.4 & $<$0.5 \cr
218 & 14  & 0.224 & 09 52 50.30 & + 7 50 34.6 & 20.9 & 2.6 & 0.9 & --- & --- \cr
219 & 45  & 1.272 & 12 54 56.70 & +56 49 41.0 & 21.9 & 2.4 & 0.8 & $<$0.5 & $<$0.6 \cr
220 & 23  & 0.193 & 17 26 19.27 & +74 48 01.9 & 19.9 & 3.8 & 1.3 & $<$0.3 & $<$0.7 \cr
223 & 17  & 0.288 & 16 33 09.60 & +57 10 40.0 & 19.3 & 2.1 & 1.2 & --- & --- \cr
232 & 16  & 0.227 & 10 08 58.82 & +50 37 30.7 & 19.9 & 2.2 & 1.2 & $<$0.3 & $<$0.6 \cr
248 & 2   & 0.274 & 09 09 43.56 & +43 02 54.6 & 21.9 & 2.4 & 0.9 & $<$0.5 & $<$0.6 \cr
255 & 7   & 0.260 & 07 59 06.74 & +37 32 35.9 & 21.2 & 2.6 & 1.2 & $<$0.4 & $<$0.7 \cr
258 & 1   & 0.698 & 11 17 50.71 & +07 57 11.9 & 21.5 & 2.4 & 0.4 & --- & --- \cr
273 & 23  & 0.433 & 10 42 50.30 & +11 51 19.0 & 20.6 & 2.3 & 1.3 & --- & --- \cr
278 & 10  & 0.091 & 13 31 52.15 & +11 16 49.8 & 16.7 & 3.0 & 1.2 & $<$0.1 & $<$0.5 \cr
281 & 21  & 0.347 & 00 10 33.51 & +10 52 31.3 & 20.2 & 2.4 & 1.1 & $<$0.3 & $<$0.6 \cr
293 & 13  & 0.189 & 08 20 12.69 & +37 35 02.8 & 21.6 & 2.6 & 0.9 & $<$0.4 & $<$0.5 \cr
}
\tabletext{FID: RIXOS field number; SNo: RIXOS source number; {\sl z}: AGN
redshift; RA and Dec: position of the optical counterpart in J2000;
$m_{\rm V}$: apparent V band magnitude measured from the optical spectra; 
\aopt: energy index of the best-fitting power-law to the optical continuum;
\aox: energy index of slope connecting 5000~\AA\ to 2~\keV; \aro: energy index
of slope connecting 1.4~GHz to 5000~\AA; \arx: energy index of slope connecting
1.4GHz to 2~\keV.}
\endtable

\subsection{Analysis of the X-ray data}

As part of the original survey analysis, the PSPC data for all AGN were divided
into three bands [0.1 to 0.4~\keV\  (channels 8 to 41); 0.5 to 0.9~\keV\
(channels 52 to 90) and 0.9 to 2.0~\keV\  (channels 91 to 201)] and these were
combined to produce `spectra' with three data points for each source. The
spectra were fitted with single power-law models using the method described  in
Mittaz \etal (1997), which finds the best-fit by minimizing a
Poissonian-based statistic. In the fits, the absorbing column density was fixed
at the Galactic column (\nhgal) measured from the 21~cm survey of Stark \etal\
(1992). All instrumental effects, including vignetting, dead-time corrections
and particle contamination, were folded into the fitting process. The data were
also corrected for any counts falling outside the extraction circle. These are
subsequently referred to as the `three-colour' data.

For this study, we have also extracted the full-resolution PSPC spectra of
those sources which have a total of at least 100 net counts; seven of the 14
sources fall into this category.  Using the standard {\sc asterix} software,
the source spectra were extracted using circles with radii of up to 2.5 arcmin
(depending on the position of any adjacent sources) and a nearby source-free
region  was used for background subtraction.  Counts were binned to yield at
least 20 per energy bin, ignoring channels 1-11 (channels 1-8 for data taken
before the PSPC detector was changed in January 1991) and channels 201-256
where the response is uncertain. The spectra have been corrected for all
instrumental effects including vignetting, dead-time and particle
contamination.

\begintable*{2}
\caption{{\bf Table 2a.} Results of X-ray spectral fitting}
\halign{#\hfil                                                   
       &\hskip0.1truecm#\hfil                                    
        &\hskip0.1truecm\hfil#\hskip0.1truecm                    
        &\hskip0.1truecm\hfil#\hfil\hskip0.1truecm               
        &\hskip0.1truecm\hfil#\hfil\hskip0.2truecm               
       &\hskip0.1truecm\hfil#\hskip0.1truecm                   
       &\hskip0.1truecm\hfil#\hskip0.1truecm                   
       &\hskip0.1truecm\hfil#\hfil\hskip0.2truecm                    
       &\hskip0.2truecm\hfil#\hfil\hskip0.1truecm                   
\cr
\noalign{\smallskip}
    FID   &
    SNo   &
    \hfil{\sl z} \hfil   &
    \hfil{radius} \hfil   &
    \hfil{\nhgal} \hfil   &
    \multispan3{\hfil{Model with \nh=\nhgal} \hfil}   &
    Mittaz \etal \cr
    &
    &
    &
    \hfil{arcmin} \hfil   &
    \hfil{10$^{21}$ cm$^{-2}$} \hfil   &
    \hfil{\ax} \hfil   &
    \hfil{norm} \hfil   &
    \hfil{$\chi^2/dof$} \hfil &
    \hfil{\ax} \hfil   \cr
 \cr
\noalign{\medskip}
122 & 13  & 0.358 & 0.4 & 0.41 & -0.1$^{+0.3}_{-0.4}$ & 3.4  & 5/9   & \phantom{-}0.0$^{+0.5}_{-0.7}$ \cr 
122 & 21  & 0.376 & 2.5 & 0.41 & -0.3$^{+0.6}_{-0.7}$ & 2.7  & 5/5   & \phantom{-}0.2$^{+0.5}_{-0.7}$ \cr 
219 & 45  & 1.272 & 0.4 & 0.13 &  0.6$^{+0.2}_{-0.2}$ & 3.2  & 12/9  & \phantom{-}0.6$^{+0.3}_{-0.3}$ \cr 
223 & 17  & 0.288 & 2.5 & 0.18 &  1.5$^{+0.1}_{-0.1}$ & 6.6  & 54/49 & \phantom{-}1.4$^{+0.1}_{-0.1}$ \cr 
248 & 2   & 0.274 & 1.0 & 0.15 &  0.9$^{+0.2}_{-0.2}$ & 3.5  & 11/11 & \phantom{-}0.8$^{+0.2}_{-0.2}$ \cr 
258 & 1   & 0.698 & 2.5 & 0.34 &  0.3$^{+0.4}_{-0.6}$ & 7.2  & 6/6   & -0.2$^{+0.5}_{-0.6}$ \cr 
278 & 10  & 0.091 & 2.5 & 0.19 & -1.2$^{+0.4}_{-0.4}$ & 18.2 & 13/15 & -1.6$^{+0.6}_{-0.7}$ \cr 
}
\tabletext{FID: RIXOS field number; SNo: RIXOS source number; {\sl z}: AGN
redshift; \nhgal: Galactic absorbing column density in units of 10$^{21}$
cm$^{-2}$; radius: the radius of the circle used to extract the X-ray spectrum
in arcminutes;  \ax: energy index of the best-fitting power-law model; norm:
normalization of the best-fitting power-law in units of photon 10$^{-5}$
\keV$^{-1}$ cm$^{-2}$ s$^{-1}$ at 1~\keV;  $\chi^2/dof$: chi-squared per degrees
of freedom of the fit.
All errors
shown in the table are 90 percent.
Errors on  \nhgal\
are $\sim$10 percent.}
\endtable

\begintable{3}
\caption{Table 2b: Fits to three-colour data}
\halign{#\hfil                                                   
       &\hskip0.1truecm#\hfil                                    
        &\hskip0.1truecm\hfil#\hskip0.1truecm                    
        &\hskip0.1truecm\hfil#\hfil\hskip0.1truecm               
        &\hskip0.1truecm\hfil#\hskip0.2truecm               
       &\hskip0.1truecm\hfil#\hskip0.1truecm                   
       &\hskip0.1truecm\hfil#\hskip0.1truecm                   
       &\hskip0.1truecm\hfil#\hfil\hskip0.2truecm                    
       &\hskip0.2truecm\hfil#\hfil\hskip0.1truecm                   
\cr
    FID   &
    SNo   &
    \hfil{\sl z} \hfil   &
    \hfil{\nhgal} \hfil   &
    \multispan2{\hfil{Mittaz \etal} \hfil} 
 \cr
    &
    &
    &
    \hfil{10$^{21}$ cm$^{-2}$} \hfil   &
    \hfil{\ax} \hfil   &
    \hfil{norm} \hfil   
 \cr
\noalign{\medskip}
218 & 14  & 0.224 & 0.30 &  1.1$^{+0.6}_{-0.5}$ & 3.8 \cr
220 & 23  & 0.193 & 0.39 & -0.2$^{+1.4}_{-3.0}$ & 2.5 \cr
232 & 16  & 0.227 & 0.08 &  0.7$^{+0.5}_{-0.6}$ & 1.4 \cr
255 & 7   & 0.260 & 0.51 &  1.3$^{+1.1}_{-2.9}$ & 1.6 \cr
273 & 23  & 0.433 & 0.28 &  2.3$^{u}$\hskip0.5truecm & 1.3 \cr
281 & 21  & 0.347 & 0.58 &  0.7$^{+0.9}_{-1.4}$ & 3.7 \cr
293 & 13  & 0.189 & 0.46 &  0.3$^{+1.1}_{-1.9}$ & 3.3 \cr
}
\tabletext{Details as for Table 2a.
All errors
shown in the table are 90 percent; errors for F273\_23 are not given because
the fits to the three-colour data were poorly determined. Errors on  \nhgal\
are $\sim$10 percent.}
\endtable

The reduced PSPC data were fitted using the {\sc xspec} spectral fitting
software. The Galactic column density in the direction of each source (\nhgal)
was calculated by interpolating between the 21~cm measurements of Stark \etal
(1992). A single power-law model was fitted to each spectrum with a cold
absorbing column fixed at the Galactic value, allowing the index and
normalization of the power-law to be free parameters. In all cases, these
power-law models provided a good fit to the data with $\chi_\nu^2\sim$1 (see
Table 2a). 

A second fit was performed with an additional cold absorption column which was
redshifted into the rest-frame of the quasar and whose column density (\nhint)
was allowed to be a free parameter (\ie giving 3 free parameters). However, no
significant improvements to the fits (according to the F-test) were found to
any of the AGN and only 90 percent upper limits could be derived on \nhint\
(these are given in Table 3). The constraints on two AGN, F223\_17 and F248\_2,
are relatively tight however, with 90 percent upper limits of 2$\times10^{20}$
cm$^{-2}$ and 3$\times10^{20}$ cm$^{-2}$ respectively, indicating low levels of
cold gas absorption in these objects.

For those AGN with too few counts for full spectral fitting, the fits to the
three-colour data (with 90 percent errors) from Mittaz \etal (1997) are
given in Table 2b. Slopes derived from the three-colour data are also listed
for the objects with full-resolution PSPC data for comparison (see Table
2a); the two sets of slopes are consistent within the 90 per cent errors.

\subsection{Radio fluxes}

We have searched for evidence of radio activity in the red AGN using the 
NRAO/VLA Sky Survey (NVSS; Condon \etal, in preparation). The NVSS  covers the
sky north of --40$^\circ$ at a frequency of 1.4~GHz to a limiting peak source
brightness of about 2.5 mJy. At the time of writing, sky coverage was available
for ten of the 14 AGN, although none were detected. Using 2.5~mJy as an upper
limit on the radio brightness, we have calculated upper limits on the
radio-to-optical and radio-to-X-ray indices (\aro\ and \arx\ respectively) for
these ten sources, and these are listed in Table 1.

\section{Analysis and Results}

The optical-to-X-ray continua of the red RIXOS AGN are shown in Figure 2 to
illustrate the X-ray spectra relative to the optical, and the overall shape for
each individual AGN.  All of the AGN in this study have a Galactic \nh\ which
is relatively low and does not itself significantly modify the intrinsic
spectrum. The continuum may look red in the optical for several reasons: {\sl
1.} dust absorption, which is wavelength-dependent and preferentially removes
the blue photons; {\sl 2.} a contribution from the host galaxy, which is
strongest in the red part of the spectrum; and {\sl 3.} an intrinsically red
continuum produced by the nucleus and unmodified along our line of sight.
Sources which do not show evidence for absorption or a strong galactic
contribution are presumed to have an intrinsically red optical slope.

\subsection{Absorbed sources}

Supporting evidence for absorption, apart from the optical continuum slope, 
may be found from two main sources, the \Ha/\Hb\ flux ratio (\ie the Balmer
decrement) and the level of cold gas absorption local to the AGN.  If case B
recombination is appropriate for the BLRs in AGN, then the Balmer decrement in
{\sl un}absorbed AGN should be $\sim$3. Thus we first look for a high \Ha/\Hb\
flux ratio as evidence of dust absorption. 

\beginfigure*{2} 
\psfig{figure=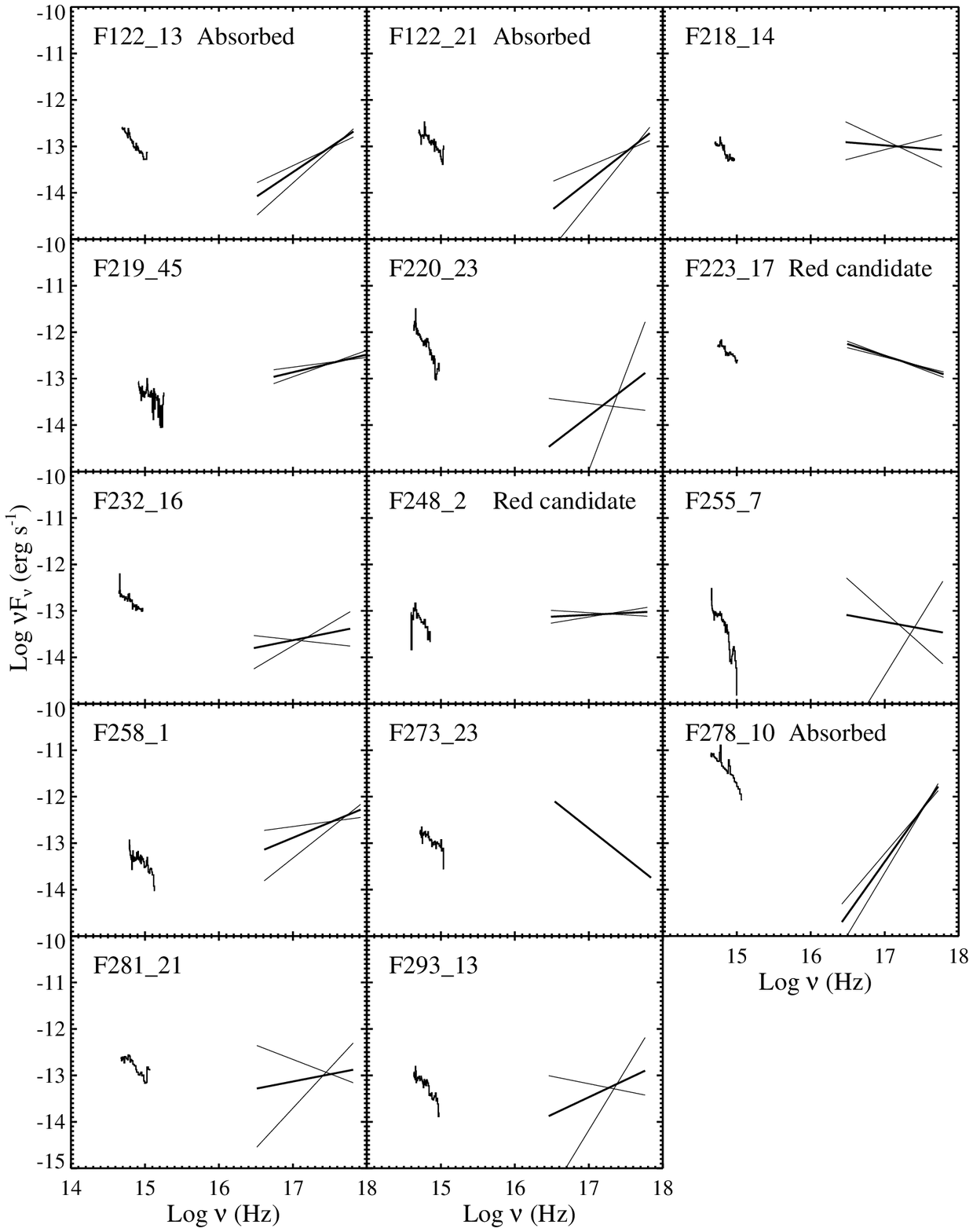,height=8.5in,width=6.5in,angle=0}
\caption{{\bf Figure 2.} Multiwavelength spectra for the red AGN. All spectra
are plotted in the AGN rest-frames.}
\endfigure

The effects of absorption may also be seen in soft X-rays. The optical
continuum can only be reddened by optically-thin dust dominated by small grains
(Laor \& Draine 1993), and since this can only exist at relatively large
distances from the centre, we assume that the dust probably resides in cold
gas. Therefore a measurement of a significant cold gas column from the PSPC
data would also suggest that a large dust column may be present, although the
presence of dust-free gas cannot be ruled out. Dust in warm gas has been
inferred in some nearby Seyferts (Brandt \etal 1996; Komossa \& Fink 1997;
Reynolds \etal 1997), however the resolution and overall quality of the PSPC
data used for the red RIXOS AGN are not sufficient to distinguish possible
columns of warm gas.

\subsubsection{Balmer decrement}

The Balmer decrement has been measured for three AGN and upper limits have been
calculated for a further five.  Errors on the Balmer decrements have been
derived by estimating upper and lower limits on the Balmer line fluxes, taking
into account the placement of the continuum and other effects such as line
blending and the relative contribution of the narrow component. The results are
listed in Table 3 and show that for all but two AGN, F232\_16 and F281\_21, the
Balmer decrement is high, suggesting significant amounts of intrinsic dust
absorption. The degree of extinction by dust, parametrized by E(B-V), has been
calculated from these Balmer decrements using the reddening curve of Cardelli,
Clayton \& Mathis (1989), and these are also given in Table 3 (column 5).

\begintable*{4}
\caption{{\bf Table 3.} Absorbing columns and intrinsic optical slopes}
\halign{#\hfil                                                   
       &\quad\hskip0.1truecm#\hfil                                    
        &\quad\hskip0.1truecm\quad\hfil#\hfil\quad\hskip0.1truecm     
        &\hskip0.1truecm\hfil#\hfil\hskip0.05truecm                   
        &\hskip0.1truecm\hfil#\hfil\quad\hskip0.05truecm                   
        &\hskip0.02truecm\hfil#                                   
       &\hskip0.1truecm\hfil#\hfil\hskip0.1truecm                    
\cr
\noalign{\bigskip}
    FID   &
    SNo   &
    \hfil{cold gas} \hfil   &
    \multispan2{\hfil dust extinction\hfil} &
    \multispan2{\hfil intrinsic \aopt \hfil}
 \cr
       &
       &
    \hfil \nhint\ 10$^{21}$ cm$^{-2}$ \hfil &
    \Ha/\Hb\  &
    E(B-V)  &
    \hfil(\nhint)\hfil &
    \hfil(\Ha/\Hb)\hfil   
 \cr
         {\sl (1)}
       & {\sl (2)}
       & {\sl (3)}
       & {\sl (4)}
       & {\sl (5)}
       & {\sl\hfil (6)\hfil}
       & {\sl (7)}
 \cr
\noalign{\medskip}
122 &  13 & $<$7.6 &             &        & $>$-1.7 &         \cr
122 &  21 & $<$25  &             &        & $>$-10  &         \cr
218 &  14 &        & 9 (2,17)    &    0.9 (0,1.4)  &   &  -0.5 (-2.7,2.6) \cr
219 &  45 & $<$3.6 &             &        & $>$-0.4 &         \cr
220 &  23 &        & $>$4        &    $>$0.1  &           & $<$3.5   \cr
223 &  17 & $<$0.2 &             &        & $>$2.0  &         \cr
232 &  16 &        & 4 (2,11)    &    0.2 (0,1.1)  &  &   1.3 (-2.0,2.2) \cr
248 &   2 & $<$0.3 & $>$2.3      & $>$0.0 & $>$2.2  &  $<$2.4 \cr
255 &   7 &        & $>$6.1      & $>$0.6 &         &  $<$0.5 \cr
258 &   1 & $<$14  &             &        & $>$-7.1 &         \cr
278 &  10 & $<$18  & $>$7        &    $>$0.7  & $>$-7.2 &   $<$-0.2  \cr
281 &  21 &        & 2 (1,26)    & 0 (0,1.8) &        &  $<$2.4 \cr
293 &  13 &        & $>$1.2      & $>$0.0 &           &  $<$2.6 \cr
}


\tabletext{{\sl (1)}: RIXOS field number; {\sl (2)}: RIXOS source number;  {\sl
(3)}: 90 percent upper limit on the 
intrinsic cold absorption column density from X-ray  spectra
in units of 10$^{21}$ sm$^{-2}$;
{\sl (4)}: Balmer decrement measured from the optical spectrum (lower and upper
limits); {\sl
(5)} dust extinction implied by Balmer decrement using reddening curve of
Cardelli \etal\ (1989) (lower and upper limits);  
{\sl (6)} hard (lower) limit on the 
energy index of the intrinsic optical
continuum derived from 
90 percent upper limits on the X-ray absorbing column \nhint\ 
(Section 3.1.2);
{\sl (7)} energy index (errors in brackets) 
of the intrinsic optical continuum
calculated using the Balmer decrement (Section 3.1.1).}
\endtable

If it can be assumed that differences in the Balmer decrement are entirely due
to reddening by dust, then it should be possible to use the \Ha/\Hb\ flux ratio
to predict the intrinsic (\ie unreddened) optical continuum slope using
reddening curves which have been derived for the dust in our Galaxy. This was
done for the red AGN by first de-reddening the spectrum using the Cardelli
\etal (1989) data and then fitting a power-law to the resultant spectrum; the
derived intrinsic \aopt\ (with errors calculated from the errors on the Balmer
decrement) is given in column 7 of Table 3. 

The table shows that due to the low signal-to-noise of the RIXOS optical
spectra and subsequently the large errors on the Balmer decrement, values for
the intrinsic (\ie the dereddened) \aopt\ are poorly constrained and in most
cases, no firm conclusions can be drawn. Sources F278\_10 and F255\_7 have
dereddened optical spectra which are no longer `red'  (the dereddened \aopt\ is
significantly lower than 2), \ie  their intrinsic optical continua are
relatively hard and typical of UV-selected quasars (under the assumptions of
this method). All other objects have upper limits which are greater than 2, \ie
their intrinsic optical continua may be red. None of the AGN have a
dereddened \aopt\ which remains significantly red (\ie \aopt$>$2).

\subsubsection{Cold gas columns from X-ray spectra}

Fits to the full-resolution {\sl ROSAT} PSPC spectra show that for two out
of the seven objects, F223\_17 and F248\_2, the best-fit \nhint\ was low
(with 90 percent upper limits of 2 and $3\times10^{20}$ cm$^{-2}$),
implying very little cold gas absorption in these AGN (although the
possibility of {\sl warm} absorption cannot be ruled out; see Section 3.1).
Column densities on the remaining five AGN are poorly constrained; intrinsic
cold gas column densities of up to 3$\times10^{22}$ cm$^{-2}$ are possible
(90 percent upper limits).

While they do not provide measurements of any cold gas column, the upper limits
on \nhint\ can be used to assess whether sources are likely to be intrinsically
red.  By assuming that any cold absorbing gas carries a given amount of dust,
the upper limits on  \nhint\ may be used as an indication of upper limits on
the amount of  {\sl dust} along the line of sight.  The intrinsic \aopt\ may
then be derived using a method similar to that described in section 3.1.1.
Limits on E(B-V) have been calculated from the limits on \nhint\ (Table 3)
using a Galactic dust-to-gas ratio, where  an E(B-V) of 1 corresponds to
\nh=6$\times10^{21}$ cm$^{-2}$ (Ryter, Cesarsky \& Audouze 1975; Gorenstein
1975). This dust extinction was then used  to calculate a {\sl lower} limit on
the intrinsic slope of the optical continuum.

The results are listed in Table 3 and show that  in most cases, the lower
limits on the de-reddened \aopt\ are  very low and little useful information
can be derived. However, two of the seven AGN have optical continua which do
remain red, F223\_17 and F248\_2, thus we identify these as `red candidate'
AGN, \ie those whose optical continua does not appear to have been
significantly modified by dust or the host galaxy.

\subsubsection{Galactic contamination}

No features similar to that of a typical underlying host galaxy (\eg Ca H and K
lines, and NaD at 5893~\AA) could be found in any of the low-$z$ AGN (\ie with
$z\le$0.5). The optical spectra of AGN at higher redshifts cover shorter
wavelengths in the rest-frame where any galactic contribution is relatively
weak. Thus the host galaxy is unlikely to cause the reddening of the optical
continuum of objects in this sample.

\beginfigure*{3} 
\psfig{figure=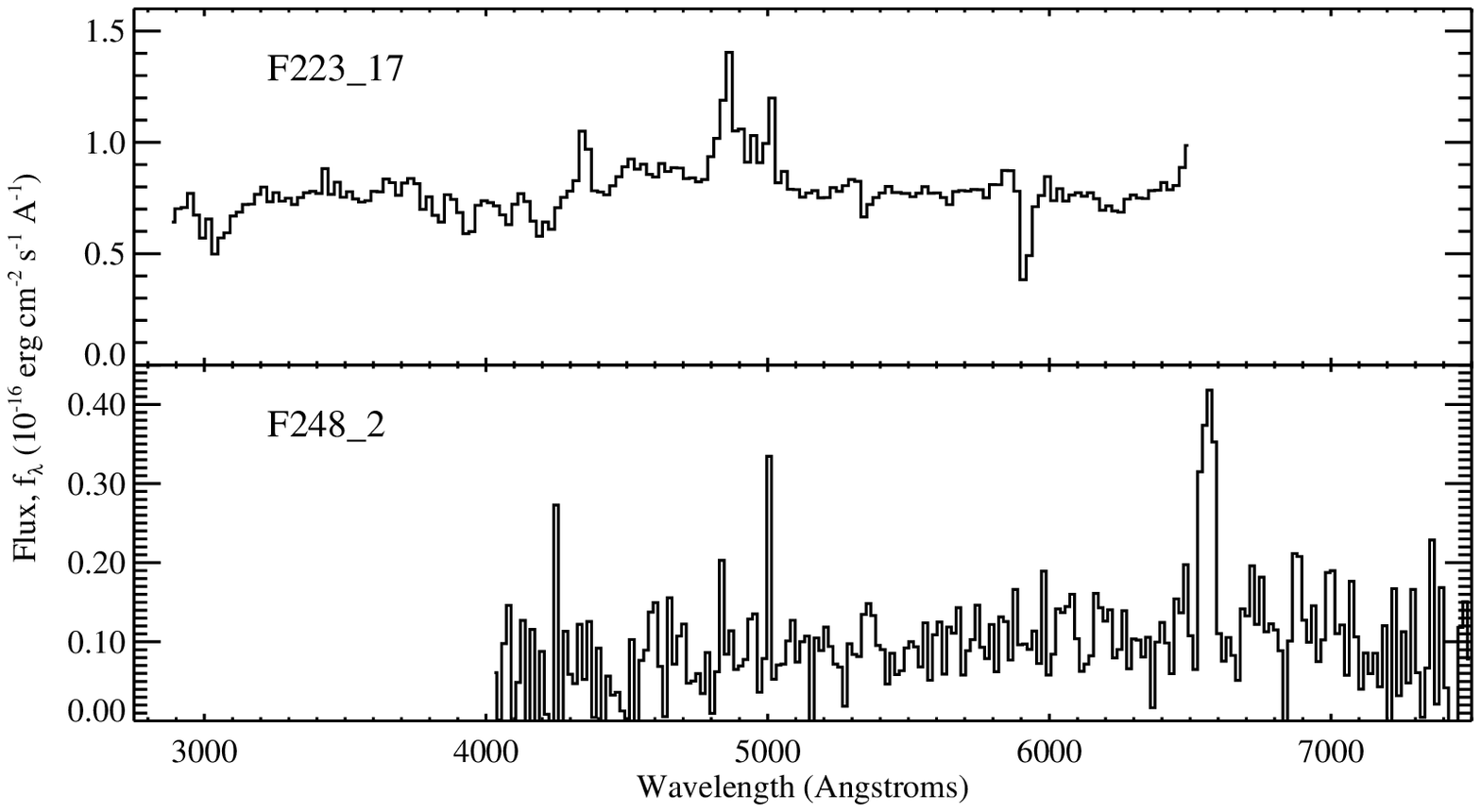,height=3.5in,width=6.5in,angle=0}
\caption{{\bf Figure 3.} The optical spectra for the red AGN
candidates (see Section 3.2). Both 
spectra are plotted in the rest-frames of the
AGN and have been binned up to a resolution of $\sim$20~\AA\ per bin
for clarity.}
\endfigure

\subsection{Red candidates}

We find that, of the 14 RIXOS AGN with an \aopt$>$2, two (the `red
candidates'; F223\_17 and F248\_2) have an optical continuum which
remains red by this definition, even after correction for 
any dust absorption by
the methods outlined in Section 3.1. Optical spectra of the red candidate
AGN are plotted in Figure 3.

The red candidates are a particularly interesting group because of the
implications they present for models of a possible underlying optical
power-law continuum in AGN, as well as for the use of optical/UV-selected
samples as representative populations of Seyferts and quasars. Thus, to
search for any evidence of other systematic differences which might provide
further clues to their physical nature, we compared their continuum
(luminosities at 5000~\AA, 2500~\AA\ and 2~\keV\ from Paper I) and line
emission properties (\Ha, \Hb\ and \oiii\ equivalent widths and full widths
at half maximum from Paper II) with the rest of the RIXOS AGN. However, we
could find no significant differences between the properties of the red
candidates and the parent sample.

\subsection{Absorbed AGN}

For the remaining 12 AGN, some degree of absorption by dust is a  possible
explanation for the reddening of the optical continuum. However, quantifying
the dust column is not a straightforward task. In Section 3.1, we showed that
the dust and cold gas columns derived  from the available data are poorly
constrained in many cases, due to the relatively poor quality of the optical
and {\sl ROSAT}-PSPC spectra.

Nonetheless, an indication of whether significant absorption might be occurring
in a source can be derived from the slope of the X-ray spectrum which was
calculated assuming a fixed Galactic \nh\ (\ie \nh=\nhgal; see, eg., Fig. 2).
This slope describes the shape of the AGN's X-ray spectrum extrinsic to our
Galaxy, and reflects how the nuclear continuum may have been modified by gas
along the line of sight. Studies of Seyferts and quasars (\eg Walter \& Fink
1993; Laor \etal\ 1997; Boyle, Wilkes \& Elvis 1997) have shown that intrinsic
slopes in the PSPC band, \ax, range between $\sim$1.3 and 1.6, and for RIXOS,
the mean \ax$\sim$1 (Paper I; Mittaz \etal 1997); at higher energies the
typical \ax\ is also $\sim$1 (\eg Comastri \etal 1992). Thus we assume that any
AGN with a relatively hard \ax, \ie \ax$<$0.5, suffers cold gas absorption
external to our Galaxy. Adopting this limit and taking into account the 90
percent errors, we find that three of the red RIXOS AGN are likely to be
absorbed by cold gas (F122\_13, F122\_21 and F278\_10), thus their optical
continua have probably been absorbed by dust.

\section{Discussion}

The very existence of AGN with red IR/optical/UV slopes has an impact on many
areas of AGN research. The proliferation of BBBs in optical and UV spectra has
naturally led to the presumption that most, if not all, AGN have such a
component. Looking for sources with a strong blue excess in optical wavelengths
is a common method used for identifying AGN; by using the presence of a blue
optical continuum as an initial defining criterion, the apparent domination of
blue sources is maintained.  However, the models of AGN structure which are
currently popular have significant covering factors of dust and gas, thus large
numbers of reddened sources should be expected. Furthermore, EUV-selected AGN
with red optical continua already present strong evidence for intrinsically red
objects. We discuss our analysis of these X-ray selected red AGN, the
implications of a large `hidden' population of reddened quasars and the
possibility of a new class of `intrinsically red' Seyferts and quasars. 

\subsection{Radio properties of red AGN}

Most of the red quasars identified previous to this study, have been discovered
in radio samples and thus most (if not all) have been radio-loud. This has led
to speculation that radio loudness may in fact be a requirement for  a red
quasar (Elvis \etal 1994b).  Of the ten sources in the RIXOS red AGN sample
for which radio data were available, none were found to be significant radio
emitters. We thus conclude that radio loudness is not a necessary condition for
red AGN. It is more likely that the apparent trend for the red quasars to be
strong radio sources is due to their selection from radio surveys.

\subsection{The effects of absorption}

Along our line of sight to any quasar nucleus, there are many possible sites
for dust (external to our Galaxy), \eg in the outer regions of the nucleus
itself, in the quasar's host galaxy, and in any intervening galaxies. The
covering factor of the dusty molecular torus alone is estimated to be $\sim$0.4
(Krolik \& Begelman 1986). It might seem surprising then, that we have a
relatively `clean' line of sight to so many AGN and that the overall population
is {\sl not} dominated by reddened sources.  Webster \etal (1995) have made an
estimate of the `missing' red sources and find that as many as 80 percent of
radio-quiet AGN may be overlooked in optical/UV-selected samples. 

Assuming that the dust resides in cold gas,  then soft X-ray-based samples will
suffer from similar selection problems   because the soft X-rays are very
readily absorbed by the gas, especially below $\sim$0.5~\keV. The RIXOS AGN
were selected by their 0.5-2~\keV\ flux to a  limit of 3$\times10^{-14}$ erg
cm$^{-2}$ s$^{-1}$. These criteria have allowed AGN with a greater range of
absorbing columns at fainter fluxes to be detected which is beneficial when
searching for objects which moderate amounts of reddening (\nhint\ of up to a
few times 10$^{21}$ cm$^{-2}$).  In Paper I, we placed a lower limit of
$\sim$0.3 on the fraction of AGN which have significant gas and dust columns
(\ie with \nhint$\ga10^{21}$ cm$^{-2}$), \ie we find that {\sl at least} 
a third of AGN have moderate to large absorbing columns.

In this paper, we have highlighted the reddest of the RIXOS AGN, \ie those with
an \aopt$>$2. Fourteen of the 160 RIXOS AGN meet this criterion and of these,
three show positive evidence of absorption, based on the slope of the
X-ray spectrum extrinsic to our Galaxy (see, eg., Figure 2). Modelling of the
X-ray spectra has shown that equivalent intrinsic cold gas columns of up to
10$^{22}$ cm$^{-2}$ are possible. This is small compared to the expected column
density of the dusty molecular torus however ($\sim10^{24}$ cm$^{-2}$; Krolik
\& Begelman 1988); if such large columns are typical of AGN in general then it
should indeed be anticipated that many AGN are being missed.

The consequences of using a select group of relatively unabsorbed  objects to
represent the properties  of a much larger, mixed population of AGN are
difficult to assess. If, as suggested by the `unified model' (see \eg Antonucci
1993 for a review), the observed properties of an AGN are largely determined by
the orientation of the molecular torus, this presumption is valid {\sl unless}
there are other angular-dependent properties in the inner regions. For example,
if emission from an accretion disc is angular dependent  (\eg Sun \& Malkan
1989; Czerny \& Elvis 1987; Madau 1988) and the disc is co-aligned with the
torus, only the disc's properties at relatively face-on angles can be observed.
In Paper II however, we found that angular-dependence may not accurately
reflect the amount of dust present, in which case there may be more fundamental
implications for the `hidden' population, \eg the amount of obscuring  dust and
gas may be related to the intrinsic source luminosity, black hole mass and/or
host galaxy type; it is only possible to speculate at this stage.  

\subsection{Unabsorbed red AGN}

Out of the 14 AGN studied, two show convincing evidence for little or no
absorption and are probably `intrinsically' red. These objects have  low cold
gas columns  and in general their X-ray emission lies well above an
extrapolation of the optical continuum. There is {\sl no} BBB emission in these
sources down to wavelengths of $\sim$3000~\AA; other reports of AGN with weak
optical BBBs have also been made (\eg McDowell \etal 1989). The intrinsically
red RIXOS AGN may be similar to some EUV-bright Seyferts whose BBBs are so hot
that they are not observed even in the UV, but are unusually strong in the EUV
and soft X-rays, even showing evidence of a high-energy turnover at
$\sim$1~\keV\ (\eg Puchnarewicz \etal 1995). F223\_17 is the most likely
analogue of these. Alternatively the BBB may be emitted away from our line of
sight, or entirely suppressed.

Another important question posed by the intrinsically red sources is the nature
of the optical continuum itself. There are two components which are known to be
strong in the optical range, the BBB and the host galaxy, yet there is evidence
for neither in these AGN. What, then, is the origin of the red, continuum
component which is observed? The resemblance of the radio to optical spectra of 
`red' or `optically-quiet' radio sources to blazars (\eg Rieke \etal 1979;
Ledden \& O'Dell 1983; Tananbaum \etal 1997) is intriguing and there may be
some relationship between the two. From a different approach, Ulvestaad \&
Antonucci (1986) examined three BL Lacs in greater detail and suggested that,
if seen from a less pole-on direction, these would probably be classified as
`optically-dull' radio galaxies. Perhaps the red quasars are blazars seen close
enough along the jet axis that the blazar continuum is observed, but not strong
enough to swamp the emission lines.  

Tananbaum \etal (1997) have suggested that there may be a physical link between
the radio emission and the relatively low optical emission in the galaxy
J~2310--43, based on the model of Donea \& Biermann (1996), where optical/UV
emission from an accretion disc is suppressed when the base of the radio jet is
large. The intrinsically red AGN  are all radio-quiet however, thus an
unusually large radio jet is unlikely to be the cause of the red optical/UV
continuum.

\section{Conclusions}

We have presented a study of 14 `red AGN' from the RIXOS sample. The sources
all have relatively soft optical continuum slopes, with an \aopt$>$2. Of
the ten for which radio data are available, all are radio-quiet. At least
two of the red AGN have very low cold gas absorbing columns
and we term these `red candidates', \ie AGN whose nuclear optical continuum
slope is very soft and shows no sign of the BBB. Of the remaining objects,
three have an \ax\ significantly harder than 0.5 and are probably absorbed.

Red AGN as a class have received little attention relative to their blue
counterparts (which are considered the norm). The most popular AGN selection
techniques (optical and soft X-ray) will have missed many of them; it has been
estimated that perhaps 80 per cent of radio-quiet AGN may have been reddened by
dust and are unaccounted for. Thus our interpretation of population studies of
AGN from predominantly unabsorbed sources are suspect at best, particularly if
dust absorption is {\sl not} closely linked with the orientation of the
nucleus.

While possible, compelling, links have been made between red quasars (which are
radio-loud) and blazars, our red candidate for which radio data is  available
is radio-quiet, making such an explanation unlikely in this case, unless this
red, `blazar-like' component has a break between the radio and the optical. For
most objects, an excess of soft X-ray (0.5-2~\keV) emission above an
extrapolation of the optical `blazar' component is also required.

A very good way of searching for red AGN in the future will be from
serendipitous detections of AGN in hard X-rays using deep fields from missions
like {\sl XMM} and {\sl AXAF}. The effects of cold gas absorption are greatly
reduced at higher X-ray energies than those sampled by, say {\sl ROSAT}, and
the vast improvements in throughput and spectral resolution promised by these
telescopes, especially by {\sl XMM}, should probe even deeper into the nucleus
than was previously possible. If indeed there are many red AGN waiting to be
discovered, the notion that `all AGN are blue' must be relaxed, at least until
such studies prove otherwise.

\section*{Acknowledgments}

We thank all in the RIXOS team for their work in obtaining and reducing the
data. We are especially grateful to Ian McHardy and Niel Brandt for their help
and advice; also to the anonymous referee whose report improved the paper. The
RIXOS project has received observing time under the International Time
Programme offered by the CCI of the Canarian Observatories and has received
financial support by the European Commission through the Access to Large-Scale
Facilities Activity of the Human Capital and Mobility Programme. This research
has made use of data obtained from the UK {\sl ROSAT} Data Archive Centre at
the Department of Physics and Astronomy, University of Leicester (LEDAS). We
also thank the Royal Society for a grant to purchase equipment essential to the
RIXOS project.

\bye